# Experimental confirmation of breach of the basic postulate of Statistical Physics

Vladimir V. Savukov

In the course of computer modeling of the most probable stationary macrostates of non-ergodic closed systems, a forecast was obtained about the existence of limits of applicability of the basic axiomatic postulate of statistical physics, known as the Principle of Equiprobability of each realizable microstate. Moreover, for such systems, thermodynamic equilibrium is no longer the only permissible stationary state. The possibility of other most probable macrostates is predicted, which are characterized by the presence of a stable anisotropy of the polarization parameters of thermal radiation filling these systems. The article presents successful results of direct experimental verification of the above prognosis on a real physical installation. Important regularities inherent in the used mathematical model are noted.

# ABSTRACT

In the course of computer modeling of the most probable stationary macrostates of non-ergodic closed systems, a forecast was obtained about the existence of limits of applicability of the basic axiomatic postulate of Statistical Physics, known as the **Principle of Equiprobability of each realizable microstate**. Moreover, for such systems, thermodynamic equilibrium is no longer the only permissible stationary state. The possibility of other most probable macrostates is predicted, which are characterized by the presence of a stable anisotropy of the polarization parameters of thermal radiation filling these systems. The article presents successful results of direct experimental verification of the above prognosis on a real physical installation. Important regularities inherent in the used mathematical model are noted.

**OCIS:** 000.6590, 260.0260, 260.5430

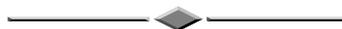

Article revision dated 2024-11-15 (click to refresh)

# TABLE OF CONTENTS

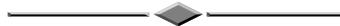

## Introduction

This paper is published as part of a research project devoted to the analysis of the limits of applicability of the axiomatic principles of Statistical Physics [1-3, 5-10, 13]. The existing methodology of Statistical Physics of equilibrium systems is based on the hypothesis of the equal probability of all microstates accessible to the closed system under consideration [1].

This means the following:

1. The most probable stationary state of a closed (isolated from the external environment) physical system is called the equilibrium state. The equilibrium state is macroscopic. It represents the totality of all microstates available to the system, i.e. those specific states, each of which can be realized at a given energy level.

2. At each fixed moment of time, the equilibrium state is materialized through one of its constituent microstates. In this case, the system can be detected with equal probability in any of the microstates forming its equilibrium macrostate.

Point 2 of this axiomatics allows us to declare how exactly the equilibrium state of a closed system should look like, and determines the directionality of stochastic processes in time. The latter means that, for example, the interaction of photons of thermal radiation inside such a system with any optical element in it is not able to change the macroscopic parameters of this radiation if they already correspond to the described definition of the equilibrium state.

Earlier [1-3] it has been suggested that there exist non-ergodic[1] quantum systems whose behavior lies outside the "zone of responsibility" of Statistical Physics. This is due to the fact that the phase trajectory of each quantum particle[2] is not continuous at the level of the impetuses subspace. The ability of quantum particles to "disappear" and "appear" in different parts of the phase space accessible to them opens the possibility of existence in this space of sources and effluents of phase trajectories having not the same density in the same local parts of the phase volume. The resulting stable in time non-zero divergence of the flux of phase trajectories in particular parts of the phase space can make this system non-ergodic and its properties inconsistent with the axiomatics of Statistical Physics.

The said gave ground to assume that under certain conditions the diffuse photon gas[3] can change the initial isotropic macrostate to the anisotropic one, which in this case will be more probable. In other words, microstates of photons of thermodynamically equilibrium radiation, initially uniformly filling some volume of phase space, can be redistributed in it due to "splitting" of phase trajectories of single photons into multiple coherent scattering channels, for example, into diffraction orders – after interaction of radiation with the grating. If the grating is an integral part of a closed physical system, the Boltzmann entropy of such a system is capable of decreasing with time. This paradox is overcome by defining entropy through Shannon's formula. This allows us to more correct-ly consider entropy as a measure of the probability of a macrostate of a closed system, without resorting to the postulate of the equal probability of its microstates, used in the definition of Boltzmann entropy [13].

---

[1] The property of ergodicity assumes the reliability of the microcanonical hypothesis of Statistical Physics about the identity of the results of time averaging and phase averaging when calculating the values of the macroscopic parameters of the system.

[2] Applied to quantum particles the notion of a phase trajectory can be preserved by its redefinition on the basis of Ehrenfest's theorem [1].

[3] By diffuse photon gas here we mean unpolarized incoherent electromagnetic radiation, for individual photons of which any angular orientation of their wave **k**-vectors in geometric space is equally probable.

Further presented the results of computer simulation of the most probable stationary macrostates of non-ergodic closed systems in which thermodynamically equilibrium Planck radiation spontaneously acquires anisotropic polarization, and also summarize the summation of field experiments on successful detection of aforecited effect on real physical installation[1].

## A system based on a phase diffraction grating and its disadvantages

It was assumed that the above-mentioned effect could be used for passive localization of objects in thermodynamic equilibrium with the environment (for example, in hidden security systems, etc.). For this purpose these objects should be "marked" by diffraction gratings, the surface of which becomes visible when observing them through a thermal imager with a polarization filter [1]. However, the described technical solution has a number of significant drawbacks.

The main problem is that the anisotropy of polarization parameters of a diffuse photon gas, arising after its interaction with a diffraction grating, is clearly manifested if this gas is monochrome. But the state of thermodynamic equilibrium is characterized by thermal radiation with a Planckian frequency distribution. In this case, the diffraction orders of scattering of photons belonging to different parts of the spectrum compensate each other to a very large extent (about 95-98%) at the level of the total energy brightness. Earlier that there was even an unconfirmed assumption that such mutual compensation necessarily reaches 100 % [1], and this would not allow to observe the predicted effect when using thermal imagers with bolometric type matrices[2].

Fig. **1** shows a series of graphic images illustrating this circumstance. Each graph is plotted in a polar coordinate system such that its center corresponds to the zero value of the reflection angle when the macroscopic surface of the diffraction grating is viewed from the outside. The magnitude of the reflection angle is proportional to the polar radius, and at the periphery of the graph the value of this angle approaches $90°$. The azimuthal angle of observation of the grating surface is determined by the polar angle.

The original light field is a diffuse radiation with a total number of photons in the statistical trial $N = 285{,}749{,}842$. Fig. **1**a, the indicatrix describes the angular brightness distribution of a monochromatic (wavelength $\lambda = 10$ μm) diffuse light field reflected from a perfectly conducting phase linear grating (step $d = 8.200$ μm, total depth of the sinusoidal microrelief profile $h = 3.116$ μm, microrelief lines oriented vertically) This graph is automatically scaled to maximally reveal all existing contrasts in the density of the scattered light flux. In Fig. **1**a, the image of the indicatrix contains only unsystematic manifestations of fluctuations, which, in accordance with Lambert's law, do not form any macroscopic gradients [1].

Fig. **1**b shows an image of the calculated probability density of the polarization angle α, defined as the arctangent of the ratio of the amplitudes of mutually orthogonal components of the electric vector in an arbitrary coordinate system [11]. This graph contains strongly pronounced gradients caused by the diffraction of monochrome radiation on a reflective grating.

---

[1] Previously, the existence of such an effect was confirmed for the monochrome radiation [2].

[2] This assumption was the reason that in the first experimental setup [2] the object of study was a pre-stochasticized monochrome photon gas, instead of real thermal radiation with Planck spectrum.

Figs. **1**c и **1**d show images, respectively, of *S*- and *P*-indicatrixes, which, according to the computer prediction, can be observed on the screen of a thermal imager equipped with a polarization filter (analyzer) – as a result of diffraction of monochrome diffuse radiation on the grating.

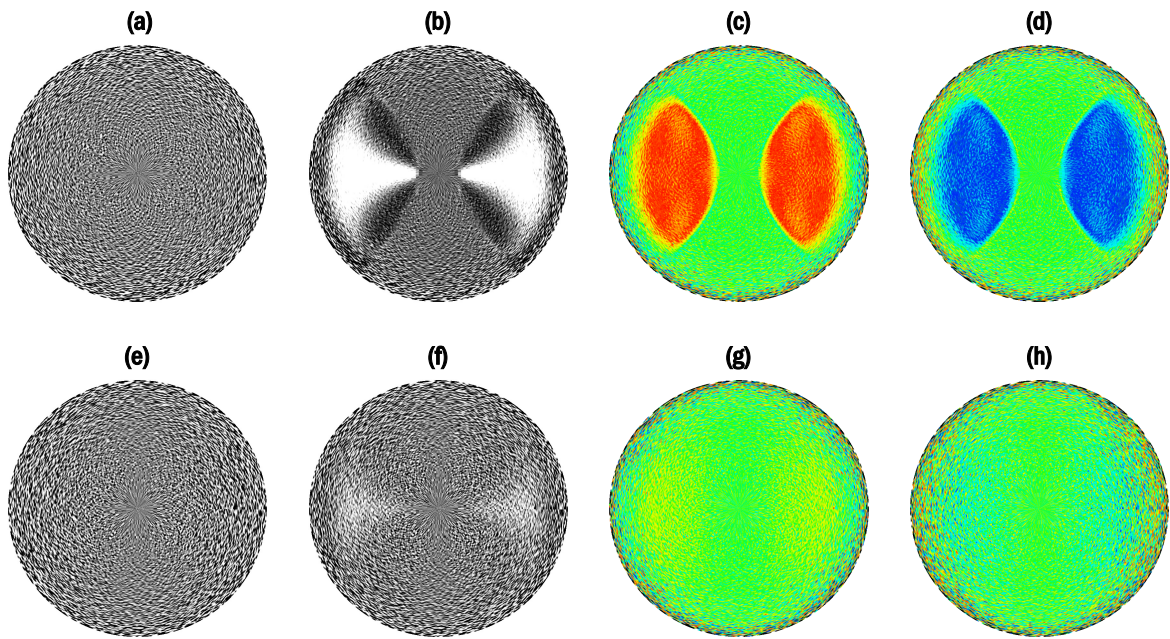

Fig. 1.  Polarization characteristics of the initially diffuse light field scattered by a phase diffraction grating for the purpose of passive location of a marked object:

–  The top row (1a-1d) contains information for the monochrome light field;

–  The bottom row (1e-1h) contains information for a light field with a Planck spectrum.

For comparison, Fig. **1** shows the second row of graphic images (**1**e, **1**f, **1**g и **1**h), containing information similar to that in the first row (see Figs. **1**a, **1**b, **1**c и **1**d), but corresponding to the case of not a monochrome, but a Planck spectrum of an isotropic photon gas, which fully meets the definition of thermodynamically equilibrium radiation with a temperature of $290 °K$.

The content of graphs **1**g and **1**h indicates that the methodology of experiment considered is practically unsuitable for working with real radiation that has an extended (Planck) spectrum. The use of a frequency narrow-band filter at the input of the thermal imager along with the analyzer (for separation of the monochrome component) is likely to lead to a fatal drop in the value of the analyzed signal, the initial level of which is already low at a typical ambient temperature of $\sim 300 °K$.

Of the other disadvantages of using diffraction gratings to polarize initially thermodynamically equilibrium radiation, two more should be mentioned:

– The nature of the signal recognized by the thermal imager strongly depends on the angle from which the grating surface is visible. There are such combinations of reflection angles and azimuth angles that form a kind of "dead zones" that exclude the detection of polarization gradients even in a monochrome radiation environment.

– Special requirements for the geometry of the microrelief of a diffraction grating can significantly increase its cost, since along with the pitch and depth of this microrelief the strictly sinusoidal shape of its profile is regulated [2]. In addition, this microrelief is very vulnerable to any external impact and can be easily damaged. The use of protective coatings on the grating surface can significantly reduce or even completely exclude the appearance of the polarization effect in question.

## Non-ergodic system based on a flat dielectric mirror

During the simulation modeling of the most probable macrostates of closed physical systems, it was discovered that diffraction polarization on regular structures is not the only mechanism that contributes to the emergence of negentropic processes in such systems. Similar possibilities have been identified in designs in which dielectric mirrors are used as optical elements instead of gratings. In this case the break of phase trajectories of particles, necessary for appearance of non-ergodic properties of the system, occurs in the process of overcoming by photons of the boundary between the internal volume of the mirror and the external medium.

Let us consider the simplest variant of a closed system with an optical element in the form of a flat dielectric mirror. A computer model of such a system, initially in a state of thermodynamic equilibrium, forecasts the spontaneous emergence of anisotropic polarization, noticeable when observing the mirror surface at an angle of reflection equal to Brewster's angle. The specified anisotropy consists in violation of equal proportions between *S*- and *P*-components of the registered radiation, which can be detected during its filtering by the analyzer.

The expected effect should be the stronger the higher the refractive index value of the dielectric mirror material. The most suitable for this purpose germanium Ge and zinc selenide ZnSe have refractive indices that, within their internal transmission windows, depend little on the radiation frequency. Thus, different frequencies of the Planck spectrum will correspond to approximately the same values of the Brewster angle[1], i.e., in contrast to polarization on a diffraction grating, here self-compensation of anisotropy manifestations will be weakly expressed.

In order to maximally match the properties of the simulation model under study to the predicted characteristics of the real physical system, it was decided to use a special way stochasticizing the photon gas parameters in the dielectric mirror thickness (see the description of the germanium optical Ge-window to Fig. **5.1**).

Fig. **2** shows a schematic diagram of a physical installation designed to identify the predicted effect (see Fig. **5** for its specific implementation).

## List of installation elements (see Fig. 2):

1. Thermostated photometric hollow ball confining a quasi-closed system[2]. The shell of the ball containing a thick layer of thermal insulation minimizes the impact of external heat fluxes on objects placed inside it.

2. The inner surface of the ball 1, covered with an absorbing material with properties close to those of a black body. This could, for example, be Vantablack™ – a special substance made of carbon nanotubes, characterized by a total integral reflection coefficient of ~ 0.045% [4].

3. Cylindrical rod capable of rotating about its axis. It is used to place the dielectric mirror 4 attached to it in the center of the ball 1. The angle at which the optical axis of the input aperture of the recording device 6 is oriented to the plane of the mirror 4 is changed by rotating the rod 3.

4. The main optical element, which is a flat dielectric mirror made of a material with a high refractive tive index and an internal transmission window of at least 8-14 μm in width (germanium Ge is most suitable). The mirror should not have anti-reflective, protective or any other coatings on its the working surface!

---

[1] Brewster's angle $\theta_{Br}$ depends on the ratio of the refractive indices of the dielectric mirror material $n$ and its surrounding medium $n_0$: $\theta_{Br} = \arctan(n/n_0)$ .

[2] The quasi-closed nature of a physical system allows an external observer to obtain information about its internal state without significantly changing this state.

5. Infrared radiation analyzer in the range of 8-14 μm. Must be able to rotate around its optical axis. Radiation components filtered for recording device 6 are determined by a combination of the transmission angle of this analyzer with the rotation angle of dielectric mirror 4 on rod 3.

6. A recording device, for example, a thermal imager with an operating range of $\lambda \approx$ 8-14 μm or a low-temperature (~300 °K) total radiation pyrometer. The thermal imaging variant provides greater demonstrativeness of the recorded information. It is better suited for obtaining visual results (see Fig. 5).

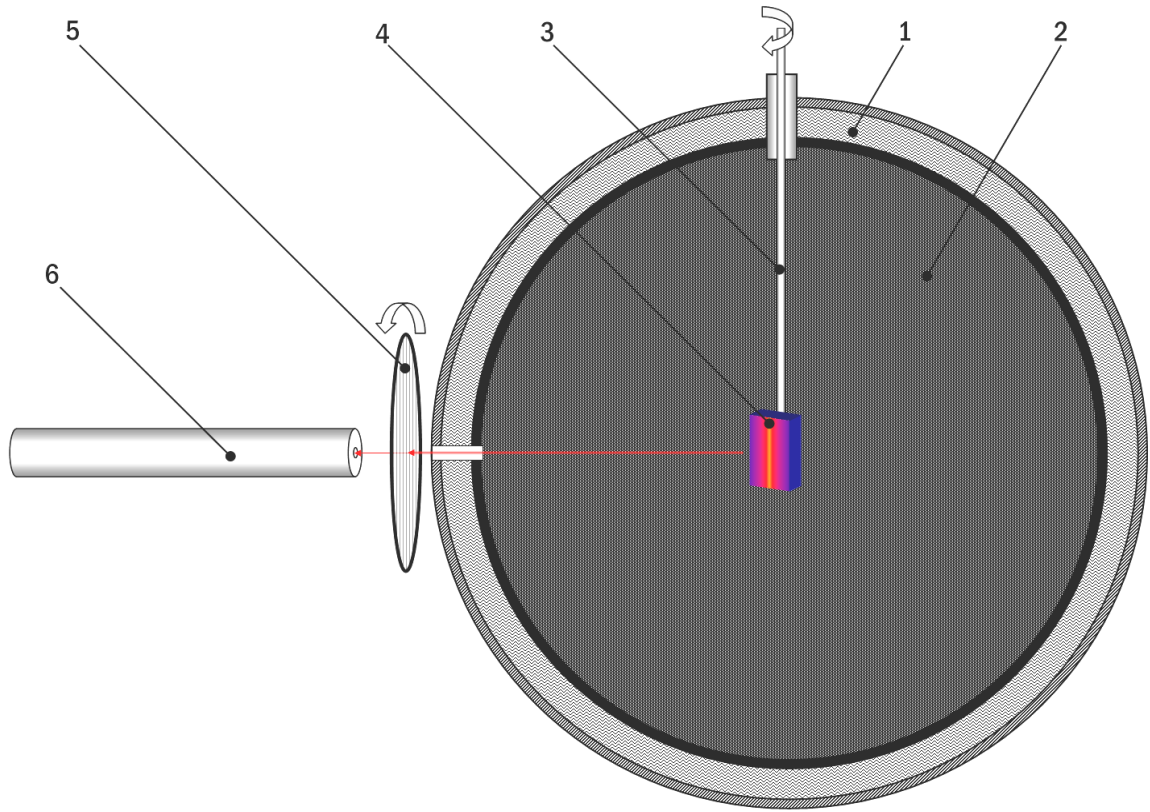

Fig. 2. Scheme of the experimental setup based on a flat dielectric mirror

This experimental setup is capable of detecting an excess of the energy brightness of the *S*-component or a corresponding dip in the brightness of the *P*-component[1] in the composition of the radiation flux recorded by device 6 during the rotation of mirror 4. These variations, according to calculations, will reach a value of ±4.0% for a mirror made of ZnSe (Brewster angle ≈ 67.4°) and ±5.6% for a mirror made of Ge (Brewster angle ≈ 76.0°). Deviations of such a significant scale correspond to such macroscopic quasitemperature gradients that can be reliably fixed even, for example, by a simple thermal imager or a low-sensitivity pyrometer.

---

[1] In the state of thermodynamic equilibrium, the relative fraction of the energy brightness of each polarization component is identically equal to ½. To make the evaluation of anisotropic radiation clear, the concept of *quasitemperature* is introduced here, for which the energy brightnesses of the *S*- and *P*-components are still conventionally considered to be identical (in fact, only the unit sum of these brightnesses is preserved).

## System with a spherical dielectric mirror

The use of a spherical dielectric mirror allows one to simultaneously observe, on sections of its surface with different azimuths, both the excess and the dip in brightness of the polarization *S*- and *P*-components contained in the thermal radiation coming from the indicated sections at Brewster's angle. A thermal imager equipped with a polarization filter should be used as a recording device. The predicted effect in this case should be equally well visible at any angle between the thermal imager and the mirror.

Figure 3 shows comparative results of modeling the interaction of a spherical germanium mirror with both monochrome and Planck diffuse radiation, and the order and semantic content of the polar graphs here correspond to those previously presented in Figure 1.

On Fig. 3a indicatrix describes the angular distribution of the brightness of a monochromatic (wavelength $\lambda = 10$ μm) diffuse light field, simultaneously fixed for all observed angles of the spherical mirror. The image of the indicatrix shows only manifestations of fluctuations that do not create any stable macroscopic gradients.

Figure 3b shows a graph of the calculated probability density of the polarization angle α. This image contains pronounced macrogradients caused by the disproportion between the polarization *S*- and *P*-components in the radiation flux coming from the side of the dielectric mirror at the Brewster angle[1] to its surface.

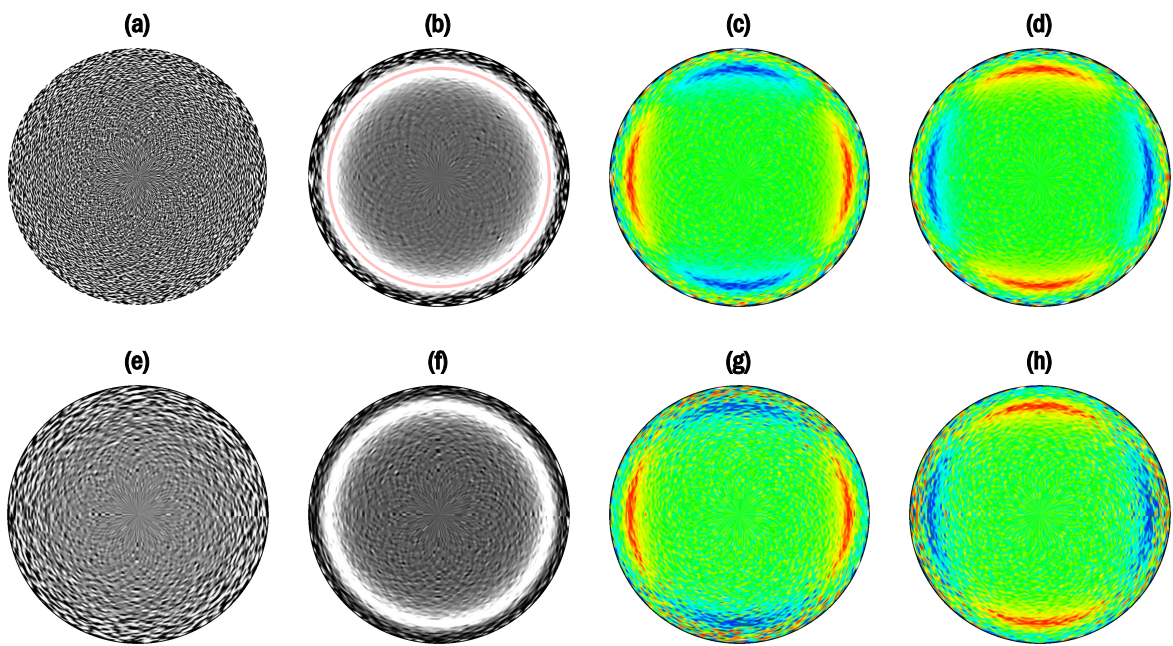

Fig. 3. Polarization characteristics similar to those shown in Fig. 1, but for the case of using a spherical dielectric mirror instead of a diffraction grating

Figs. 3c and 3d show the images of *S*- and *P*-indicatrixes, which, according to the prediction, can be observed on the screen of an analyzer-equipped thermal imager – as a result of polarization of monochromatic diffuse radiation with the assistance of a dielectric mirror.

---

[1] The angles of reflection from the dielectric mirror surface equal to the Brewster angle are marked in the polar plot of Fig. 3b with a red ring line

*Experimental confirmation of breach of the basic postulate of Statistical Physics*

For comparison, Fig. 3 show the second row of graphic images (3e, 3f, 3g and 3h), containing information similar to that shown in the first row (see Figs. 3a, 3b, 3c and 3d), but corresponding to the case of not a monochrome, but a Planck spectrum of an isotropic photon gas, appropriate to the definition of thermodynamically equilibrium radiation with a temperature of 290°K. The content of graphs 3g and 3h indicates that the methodology of the experiment using a dielectric mirror is quite suitable for working with real radiation having an extend spectrum.

The scale of manifestation of the non-entropic effect can be illustrated on the example of predicted results of application in experiments of dielectric mirrors of spherical shape made of materials with different refractive indices "$n$". Fig. 4 shows the expected images that, under previously specified conditions (equilibrium radiation, $T = 290°K$), can be observed on the screen of a thermal imager equipped with a polarization filter.

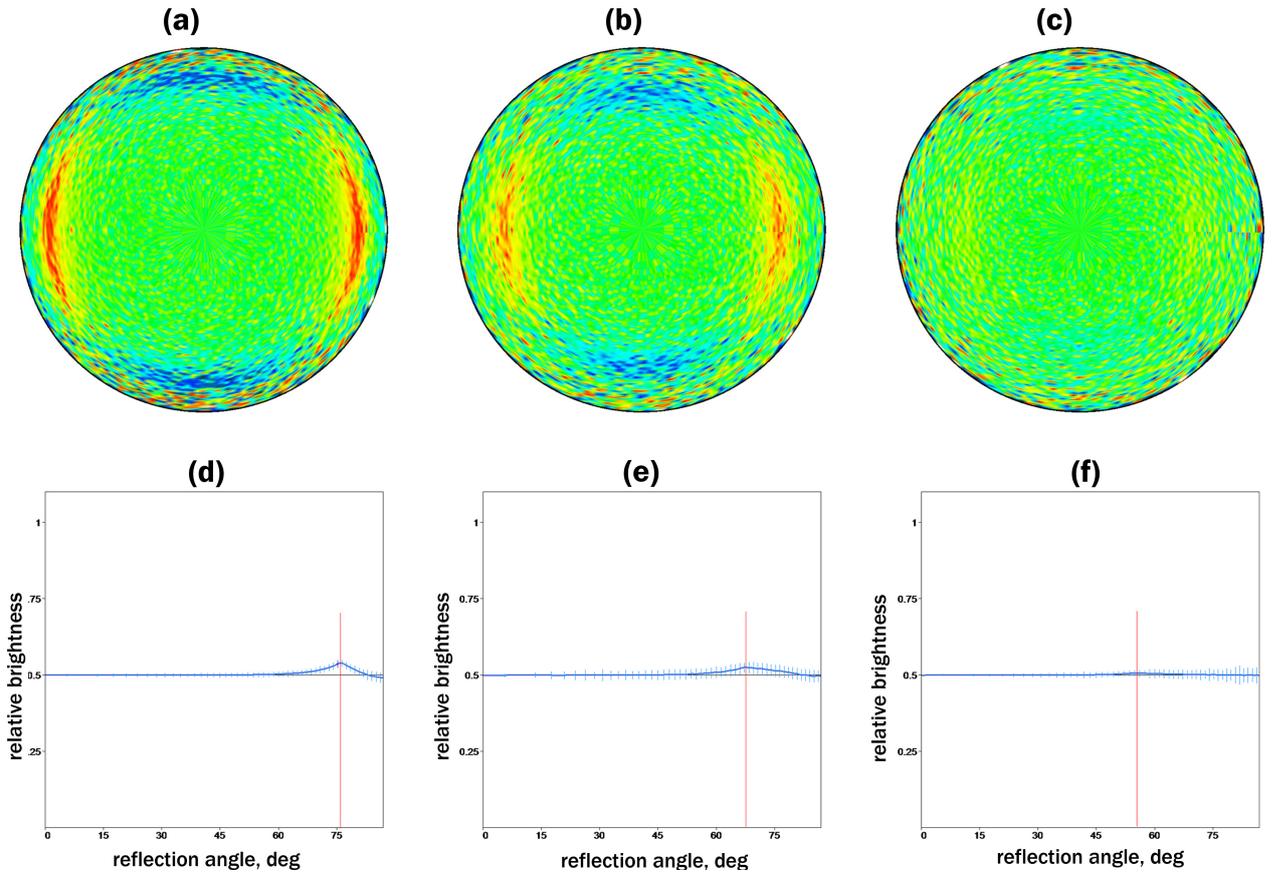

Fig. 4. Predicted images of *S*-indicatrixes for mirrors made of Ge ("a"), ZnSe ("b") and BaF$_2$ ("c"), as well as graphs of radial sections of these indicatrixes in the horizontal plane

Fig. 4a shows a prediction of the image of the *S*-component of polarized radiation going from the side of a dielectric mirror made of single-crystal germanium. Below this image is a graph (Fig. 4d) of the relative brightness of the *S*-component of that part of the emission observed at the reflection angle θ (the vertical line marks the value of the Brewster angle $\theta_{Br} \approx 75.98°$). For comparison, Figs. 4b and 4e show similar characteristics of a mirror made of polycrystalline zinc selenide ZnSe, and Figs. 4c and 4f show those of a mirror made of BaF$_2$.

The expected deviation in quasitemperature from the thermodynamic equilibrium value $T = 290°K$ recorded by the thermal imager based on the radiance of the *S*-component, will be ±5.59°K ($n \approx 4.00$) for germanium mirror, ±4.03°K ($n \approx 2.41$) for a zinc selenide mirror, and ±0.03°K ($n \approx 1.40$) for a barium fluoride mirror.

© 2024 V.V. Savukov    D. F. Ustinov "VOENMEKH" Baltic State Technical University    Page 10

If, for example, the dispersion in the **4**d graph (Ge) is considered a measure of the veracity of the mathematical expectation of the energy brightness of the S-component, then the deviation of this expectation from zero can be a consequence of fluctuation in the thermodynamic equilibrium state only with a probability of less than $10^{-4}$.

### Direct experimental verification of the existence of the predicted effect

The verification of the very fact of the existence of the expected effect was carried out using a modified scheme, namely: instead of a spherical dielectric mirror, a so-called "window" was used as the main optical element – a flat plate of optical-quality monocrystalline germanium. Fig. **5** show photographs of parts being assembled experimental setup, which is a quasi-closed physical system, the initial state of which was artificially "prepared" close to the state of thermodynamic equilibrium ($\Delta T \approx \pm 0.1 °K$).

| 1. Germanium "window" | 2. Azimuth map | 3. Analyzer "Nicole" type | 4. Infrared imager "SATIR D300" |
|---|---|---|---|

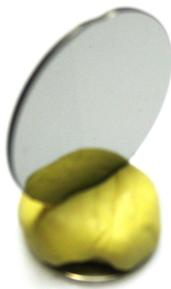 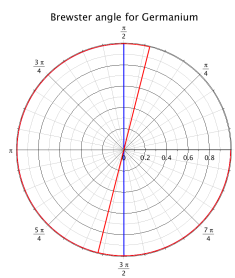 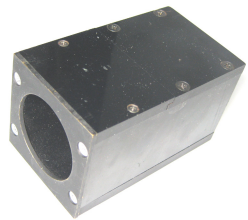 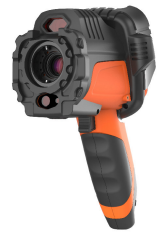

**5. Position of Ge-window "1" placed on the map "2" at Brewster's angle to the input aperture of the analyzer "3"**

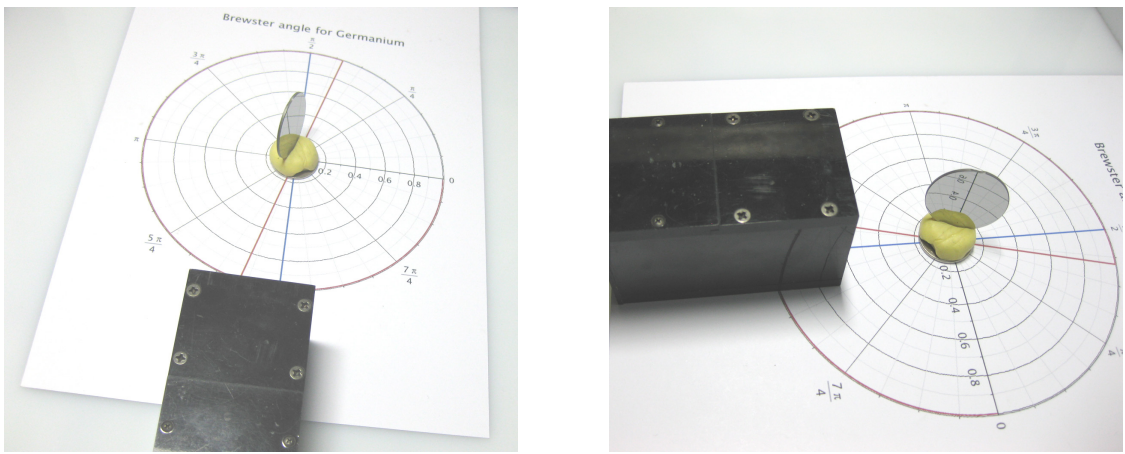

Fig. 5. Components of a setup designed to detect a predicted effect in a real physical system with a moderately uniform temperature field (ΔT ≈ ±0.1°K)

### List of installation components of the experimental setup:

1. The germanium "window" is a flat disk of optically pure germanium. Its working side is polished and does not have any antireflective or protective coatings. The reverse side of the Ge window is specially matted to create a chaotic microrelief on it. This is done in order to stochasticize the photon gas in the thickness of a given sample, which brings the parameters of the physical experiment closer to the computational model.

2. Azimuthal map is a graphical representation of the indicatrix used to set specify the angle of reflection at which the analyzer (see item 3) "sees" the surface of the Ge-window.

3. Polarized radiation analyzer of the Nicol prism type. It has such a feature as the selection of only one observed polarization component from the radiation flux. It is possible that such separation of components is the mechanism that allows, along with polarization nonreciprocity (anisotropy of isoenergetic states of closed systems), to also obtain amplitude nonreciprocity (compare the energy brightness of the aperture images in Figs. 6e and 6f).

4. The "SATIR D300" thermal imager is a device manufactured by SATIR™ Europe (Ireland).

5. Photos explaining the mutual angular arrangement of elements 1, 2 and 3 of the installation.

## Explanations of the experimental results presented in Fig. 6

Fig. 6 shows two main results of the experiment in a maximally isolated (practically closed) physical system: for example, a special shell was used to protect the installation from external radiation, etc. The outcome of each of these experiments was determined by the selected value of the reflection angle at which the input channel of the analyzer "saw" the working surface of the Ge-window:

– a reflection angle of $45°$ (see Fig. 6a) did not lead to any deviation from the isotropy of the polarization properties of Planck radiation, which characterizes the initial state of thermodynamic equilibrium of a physical system (Figs. 6c and 6e);

– the reflection angle close to the Brewster angle $\approx 76°$ for some segment of the Ge-window surface (see Fig. 6b) led to an intense luminescence of this segment (Figs. 6d and 6f), which disrupted the isotropy of the system state (see Fig. 4a, about *S*-component of the radiation).

## Explanation of the experimental results shown in Fig. 7

Carrying out the experiment in such strict version (see Fig. 6), when the system is isolated from the environment as much as possible, is a necessary step to verify the correctness of the results obtained. However, after successful completion of this check, a justified relaxation of the conditions for carrying out work with real optical elements becomes permissible.

For example, not using illumination during imaging often resulted in autofocus problems on the Ge window when it was in thermodynamic equilibrium with the environment, which was visually evident in the form of almost structureless noise (see Fig. 6c). The problem was solved by using a weak auxiliary lighting directed from the thermal imager perpendicular to the front panel of the analyzer (see Fig. 7). In this case, autofocusing was performed on frontally illuminated (and therefore clearly visible[1] against the background of stochastic fluctuations) construction elements in the internal channel of the analyzer. At the same time, the reflection angles from the Ge-window ($45°$ and $76°$) used in the experiments excluded any influence of visible light on the observed infrared image of this window. Indeed, since the germanium window is an effective dielectric mirror, then at the specified angles it reflects light frontally directed to it either across the optical axis of the analyzer (if the reflection angle was $45°$, see Fig. 6a), or in the direction opposite to the thermal imager (if the reflection angle was equal to the Brewster angle: $\approx 76°$, see Fig. 5.5 and Fig. 6b). All this helped to obtain clear images of the sought artifacts[2].

---

[1] The device features "DUO-VISION Plus" technology, which allows you to superimpose an infrared image (λ = 8-14 μm) on a visible image (λ = 0.38-0.76 μm) – for better detail of the overall picture.

[2] An *artifact* is understood here as a effect that is considered impossible (for axiomatic reasons) or extremely unlikely (fluctuations) under existing conditions. The visual manifestation of such an artifact will be a bright (5% higher than the brightness of the environment) glow of the Ge-window segments observed at a Brewster angle of $76°$.

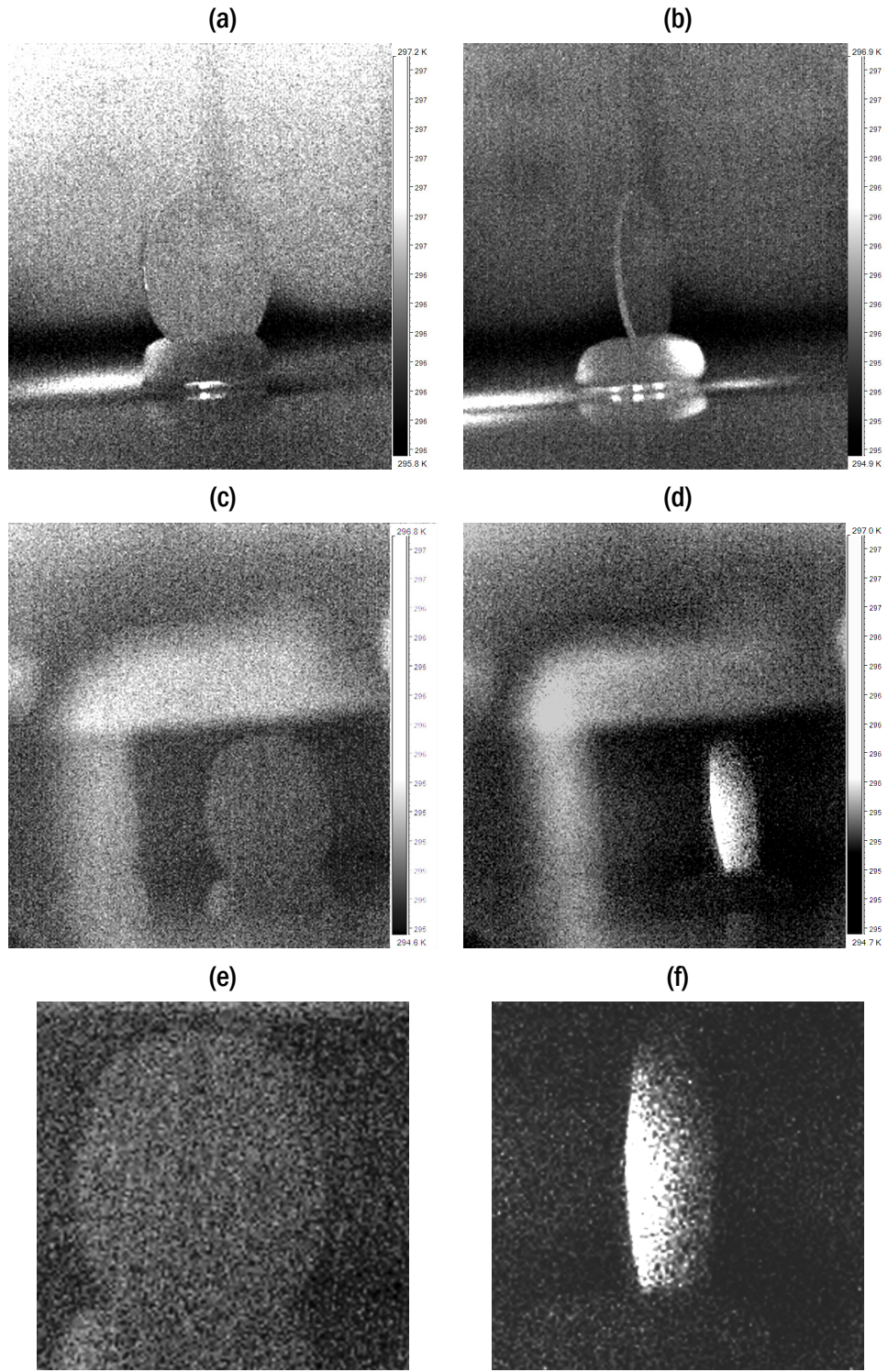

Fig. 6. Spontaneous polarization of intrinsic radiation of a quasi-closed physical system:

– a: Ge-window surface viewing angle ≈ 45˚ (no analyzer, non-infrared photo with open protective shell)
– b: Ge-window surface is seen at Brewster's angle ≈ 76˚ (no analyzer, non-infrared photo, shell open)
– c: mid-shot of the analyzer output aperture, Ge-window surface is viewing at angle 45˚ (only infrared photo)
– d: mid-shot of the analyzer output aperture, Ge-window is seen at Brewster's angle 76˚ (only infrared photo)
– e: close-up of the analyzer output aperture, Ge-window surface is viewing at angle 45˚ (only infrared photo)
– f: close-up of the analyzer output aperture, Ge-window is seen at Brewster's angle 76˚ (only infrared photo)

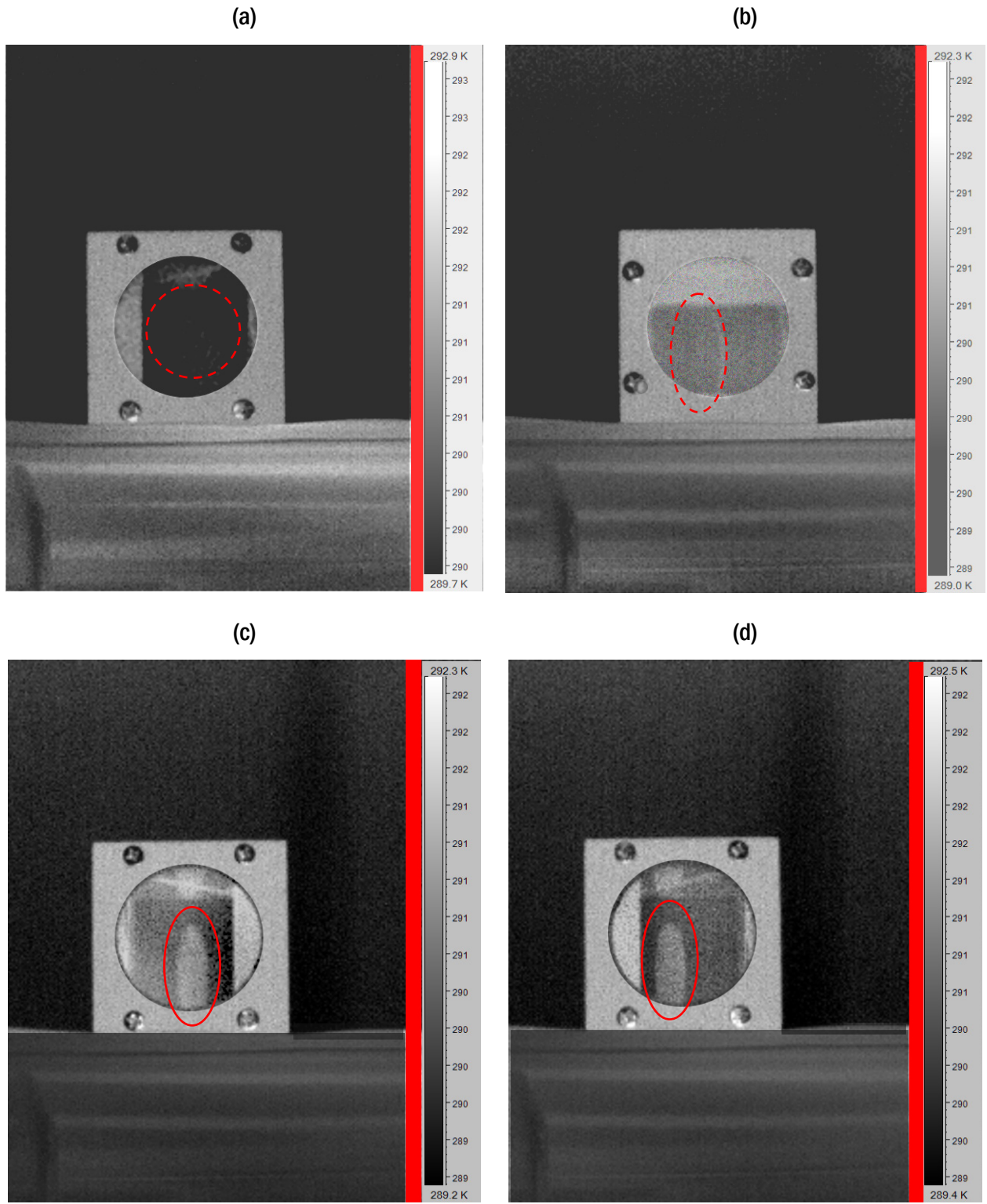

Fig. 7. Spontaneous polarization of thermodynamically quasi-equilibrium Planck radiation:

– a: Analyzer axis rotation angle: 180°, Ge-window surface is visible from a 45° perspective
– b: Analyzer axis rotation angle: 090°, Ge-window surface is visible from a 76° (Brewster's)
– c: Analyzer axis rotation angle: 000°, Ge-window surface is visible from a 76° (Brewster's)
– d: Analyzer axis rotation angle: 180°, Ge-window surface is visible from a 76° (Brewster's)

Fig. **7** show (in DUO-VISION technology) the results of four typical experiments. The outcome of each was determined by the following selected combination of two angular characteristics:

– the reflection angle at which the analyzer "sees" the working surface of the Ge-window;
– the rotation angle of the transmission plane of polarization analyzer.

a. The analyzer rotation angle[1] of 180° formally allows us to record the anisotropic polarization in the radiation coming from the optical element side. However, polarization anisotropy itself is absent here due to the fact that the angle of reflection of the studied flow from the Ge-window is assigned to be differing from the Brewster angle (45° instead of 76°).

b. The angle of reflection of the studied flow from the Ge-window is taken to be equal to the Brewster angle. However, the 90° rotation angle of the analyzer does not allow to seeing the presence of anisotropic polarization in the composition of the radiation coming from the optical element. The manifestation of this anisotropy for such an angle of rotation of the analyzer is due to the polarization component that is forcibly extinguished in this device (peculiarity of the Nicol prism). In Fig. **7**b, the red dashed line circles the expected place on the aperture where the artifact could appear visually, as if the analyzer used did not absorb one of the polarization components of the radiation passing through it[2].

c. The angle of reflection of the studied flow from the Ge-window is taken to be equal to the Brewster angle. The angle of rotation of the analyzer by 0° with respect to the transmission plane makes it possible to observe anisotropic polarization in the radiation coming from the optical element (this artifact is circled by a solid red line in Fig. **7**c).

d. The angle of reflection of ambient thermal radiation from the Ge-window is also assumed here to be equal to the Brewster angle. Therefore, turning the analyzer 180° to the transmission plane also provides the opportunity to see the anisotropic polarization pattern (the corresponding artifact is outlined in Fig. **7**d with a solid red line).

Thus, the results of all physical experiments are in agreement with the forecasts previously obtained from simulation modeling of non-ergodic systems.

## Discovered regularities of phase space parameters

Manifestations of anisotropy of photon gas, previously discovered in the analysis of the most probable stationary states of closed systems, are caused by special values of the polarization angle α. It is noted that variations of this angle cannot be arbitrary. In particular, the functional integral (**1**) calculated over the volume of the phase space accessible to such a system, always[3] has the property:

$$\int_0^{\pi/2} \cos(\alpha)^2 \, g(\alpha) \, d\alpha \equiv \frac{1}{2} \qquad (1)$$

here the non-negative function $g(\alpha)$ is the probability density of the polarization angle.

---

[1] For an analyzer, the values of the transmission and blanking angles depend strongly on both the design of the device itself and on the nature of the radiation being studied..

[2] It is possible that such amplitude nonreciprocity allows for the directed transfer of radiation energy between different parts of a thermodynamically equilibrium system.

[3] This property is not relevant if the system is not closed or far from stationarity.

For the state of thermodynamic equilibrium the function $g(\alpha)$ has the only admissible (in accordance with the axiomatics of Statistical Physics) form:

$$g(\alpha) = \sin(2\alpha) \geq 0, \quad \int_0^{\pi/2} g(\alpha)\,d\alpha = 1, \quad \forall \alpha \in [0, \pi/2) \tag{2}$$

Of course, with such a definition of probability density, it is impossible for any gradients of the quantitative ratio of the *S*- and *P*-components to arise in the system. Let us now try to introduce a modified version of the probability density function $G(\alpha)$ that would satisfy the identity (1) and the normalization (2), but still allow us to vary the polarization angle α as freely as possible:

$$G(\alpha) = g(\alpha) + R(\alpha) \geq 0, \quad \forall \alpha \in [0, \pi/2) \tag{3}$$

Let us decompose in Fourier series the function $R(\alpha)$ added into the expression of $G(\alpha)$:

$$R(\alpha) = \frac{a_0}{2} + \sum_{k=1}^{\infty}\left(b_k \sin(k\alpha) + a_m \cos(m\alpha)\right) \tag{4}$$

Then the constraints of the harmonics allowed for the function R(α) look like this:

$$\int_0^{\pi/2} G(\alpha)\,d\alpha = \int_0^{\pi/2} \sin(2\alpha) + R(\alpha)\,d\alpha \equiv 1 \tag{5}$$

$$\int_0^{\pi/2} \cos^2(\alpha) R(\alpha)\,d\alpha \equiv 0, \quad \forall a_m \tag{6}$$

Final form of the function $G(\alpha)$ taking into account conditions (5) and (6):

$$G(\alpha) = \sin(2\alpha) + \sum_{m=4,6\ldots}^{\infty} a_m \cos(m\alpha) \geq 0, \quad \forall \alpha \in [0, \pi/2) \tag{7}$$

Figure 8 shows graphs for four cases of probability densities of polarization angles. Blue lines indicate functions g(α) for isotropic equilibrium states. Circles indicate data from computer simulation models. The red lines show the approximation of these data using the function G(α), the range of acceptable values of which contains only about one quarter of all existing harmonics of the Fourier series (4). Nevertheless, the allowed harmonics (7) turn out to be quite sufficient to exactly match the modeled parameters, taking into account even the manifestations of fluctuations. This result clearly proves that the parameters of the modeled most probable stationary states of closed systems, calculated by an independent technique [3], belong to the set corresponding to condition (7).

The obtained data illustrate the earlier conclusions about the different efficiency of reflecting phase diffraction gratings and dielectric mirrors for the polarization of a diffuse photon gas with a Planck spectrum. Figures 8 show how significantly different are the probability densities G(α) generated from the primary isotropic distribution g(α) for the monochrome and Planck photon gas:

**Figure 8**a: *G* (α) - monochromatic photon gas after scattering on the reflecting phase grating

**Figure 8**b: *G* (α) - Planck photon gas after scattering on a reflecting phase grating

Anisotropic polarization of initially diffuse monochromatic photon gas and thermodynamically equilibrium radiation is effectively manifested precisely when a dielectric mirror is used for this purpose (see Figs. 8c and 8d):

**Figure 8**c: *G* (α) - monochromatic photon gas after interaction with a dielectric mirror

**Figure 8**d: *G* (α) - Planck photon gas after interaction with a dielectric mirror

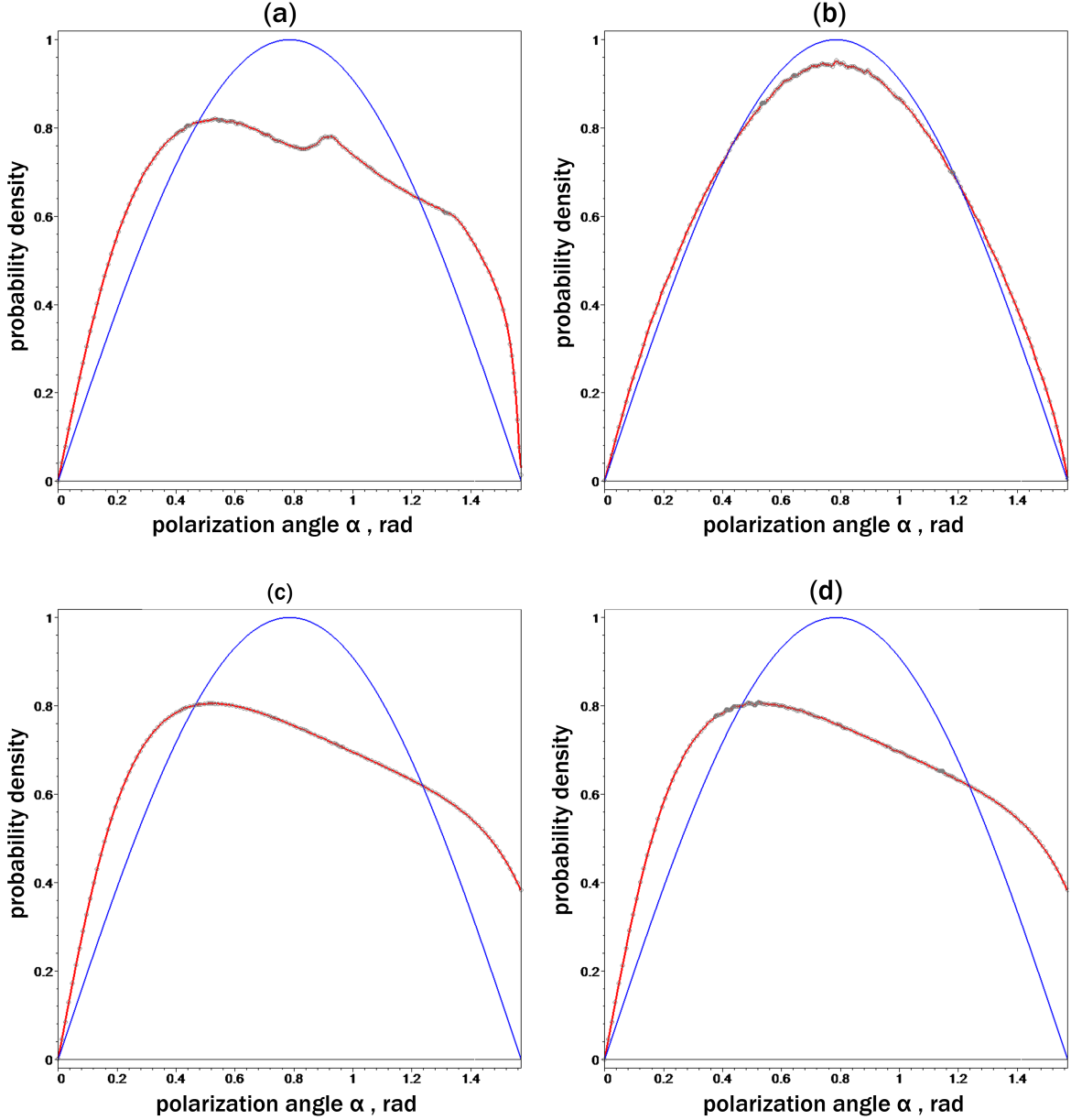

Fig. 8. Graphs of the dependence of probability densities g (α) and G (α) on the polarization angle α:

It is necessary to especially note the significance of property (**1**) for closed systems from the point of view that, despite the possibility of significant anisotropy of the polarization angle in geometric space (**7**), this anisotropy, probably, is fundamentally incapable of leading to the appearance of temperature gradients in such a system. Although, for example, for osmium Os the difference in the absorption coefficients of the *S*- and *P*-components of thermodynamically equilibrium radiation with the spectrum of a type "A" source (*T* = 2856°K) can reach almost a twofold value, no anisotropic polarization, if it appears in combination with condition (**1**), is able to change the absorption coefficient μ. This is because the dependence of this coefficient on the polarization angle is expressed by a linear parametric function in which one of the variables is the absorption coefficient μ and the other is the square of the cosine of the angle α:

$$\mu(\alpha) = \mu_1 \eta(\alpha) + \mu_0, \quad \eta(\alpha) = \cos^2(\alpha) \tag{8}$$

here $\mu_0$ and $\mu_1$ are constants that determine the properties of the material that absorbs radiation

Thus, if condition (1) is true for the phase space of a closed system, then, taking into account definition (3) and equality (6), the value of the absorption coefficient μ strictly coincides with the value characteristic of isotropic radiation - without any polarization gradients. This excludes the occurrence of temperature differences in the system, but perhaps in the future there will be ways to solve such problems on the basis of the described non-entropic phenomena.

A computer simulation model of one such method forecasts that a dielectric grating with a deep profile of microrelief and a complex refractive index of the material $ñ = n + κi$ (with a non-zero extinction coefficient $0 < κ < 0.1$), placed inside a closed system, can cause a "photon valve" effect, namely: during the relaxation process, the system will go into the most probable steady state (different from thermodynamic equilibrium), in which the relationships between the emission and absorption coefficients of radiation by the grating will not correspond to Kirchhoff's law. The resulting amplitude nonreciprocity will manifest itself in a deviation of the volume concentration of photons in the space surrounding the grating from the value determined by Planck's law, which will create a temperature difference between the grating and the radiation in contact with it.

## Conclusion

In the course of the works performed under the present project, the following result was achieved: the existence of non-ergodic closed physical systems whose most probable stationary macrostates depend on their internal organization was theoretically substantiated and experimentally confirmed. For an external observer, such deviation manifests itself in the form of stable polarization anisotropy of thermal radiation filling these systems in the above-mentioned macrostates.

Further development of this topic is to analyze the significance of the result obtained. For example, the foregoing casts doubt on the universal character of the main axiomatic principle of Statistical Physics about the equal probability of all available microstates in a closed system, which, in turn, does not exclude the revision of Boltzmann's H theorem, which is a statistical analog of the Second Law of Thermodynamics. Of course, the assessment of the prospects arising from the above should be extremely restrained. However, so far the verification of the simulation model under study has not revealed significant errors, and the reality of the predicted effects has been confirmed by the positive results of field experiments. Therefore, the continuation of search works in the accepted direction should be considered justified.

## Acknowledgment

The author is grateful to Vladimir Baloban, Head of the Scientific Research Department of the Baltic Technical University, and Alexander Tibilov, Deputy Editor-in-Chief of the Optical Journal, whose organizational assistance made the realization of this project possible.

## Funding

This research was carried out with the support of the Ministry of Education and Science of the Russian Federation (grant No. 9.1354.2014/K)

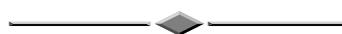